\documentstyle[multicol,aps,psfig,epsf]{revtex}
\newcommand{\la}{\left\langle}
\newcommand{\ra}{\right\rangle}
\newcommand{\be}{\begin{equation}}
\newcommand{\ee}{\end{equation}}
\newcommand{\bea}{\begin{eqnarray}}
\newcommand{\eea}{\end{eqnarray}}
\newcommand{\ba}{\begin{array}}
\newcommand{\ea}{\end{array}}
\newcommand{\piulug}{$\Pi^{u<}_{u>}$}
\newcommand{\piulbg}{$\Pi^{u<}_{b>}$}
\newcommand{\piulbl}{$\Pi^{u<}_{b<}$}
\newcommand{\piblbg}{$\Pi^{b<}_{b>}$}
\newcommand{\piblug}{$\Pi^{b<}_{u>}$}

\newcommand{\eqnMHDubk}
{\bea 
\left( -i\omega + \nu k^2 \right) u_i (\hat{k})  & 
 =  & -\frac{i}{2} P^+_{ijm}({\bf k}) \int d\hat{p} 
          [ u_j (\hat{p}) u_m (\hat{k}-\hat{p}) -
           b_j (\hat{p}) b_m (\hat{k}-\hat{p}) ] \label{eqn:udot} \\
\left( -i\omega + \lambda k^2 \right) b_i (\hat{k})  & 
 = &  - i P^-_{ijm}({\bf k}) \int d\hat{p} 
          [ u_j (\hat{p}) b_m (\hat{k}-\hat{p}) ]
       		\label{eqn:bdot}   \\
k_i u_i({\bf k}) & = & 0 \\
k_i b_i({\bf k}) & = & 0  \eea}

\begin{document}
\draft
\title{Field theoretic calculation of renormalized-viscosity,  
renormalized-resistivity, and energy fluxes of  magnetohydrodynamic turbulence}
\author{Mahendra\ K.\ Verma  \thanks{email: mkv@iitk.ac.in}}
\address{Department of Physics, Indian Institute of Technology,
Kanpur  --  208016, INDIA}
\date{May 2001}
\maketitle
\begin{abstract}

A self-consistent renormalization (RG) scheme has been applied to
nonhelical magnetohydrodynamic turbulence with zero cross helicity.
Kolmogorov's $5/3$ powerlaw has been shown to be a consistent solution
for $d \ge d_c \approx 2.2$.  For Kolmogorov's solution, both
renormalized viscosity and resistivity are positive for the whole
range of parameters.  Various cascade rate and Kolmogorov's constant
for MHD turbulence have been calculated by solving the flux equation
to the first order in perturbation series.  We find that the magnetic
energy cascades forward.  The Kolmogorov's constant for $d=3$ does not vary
significantly with $r_A$ and is found to be close to the constant for
fluid turbulence.
\end{abstract}
\vspace{1.5cm}

The statistical theory of magnetohydrodynamic (MHD) turbulence is one
of the important problems of current research.  The quantities of
interests in this area are energy spectrum, cascade rates,
intermittency exponents etc.  In this letter we analytically compute
the renormalized-viscosity, renormalized-resistivity, and cascade
rates using the field-theoretic techniques.

The incompressible MHD equation in Fourier space is given by
\eqnMHDubk
where ${\bf u}$ and ${\bf b}$ are the velocity and magnetic field
fluctuations respectively, $\nu$ and $\lambda$ are the viscosity and
the resistivity respectively, and $d$ is the space dimension.  Also,
\begin{eqnarray}
P^{+}_{ijm}({\bf k}) & = & k_j P_{im}({\bf k}) + k_m P_{ij}({\bf k});
\\ P_{im}({\bf k}) & = & \delta_{im}-\frac{k_i k_m}{k^2}; \\
P^-_{ijm}({\bf k}) & = & k_j \delta_{im} - k_m \delta_{ij}; \\
\hat{k}  =  ({\bf k},\omega); & \hspace{1cm} &
d \hat{p}  = d {\bf p} d \omega/(2 \pi)^{d+1}.
\end{eqnarray}

The energy spectra for MHD, $E^{u}(k)$ and $E^b(k)$, are still under
debate.  Kraichnan \cite{Krai65} and Irosnikov \cite{Iros} first gave
phenomenology of steady-state, homogeneous, and isotropic MHD
turbulence, and proposed that the spectra is proportional to
$k^{-3/2}$.  Later Marsch \cite{Mars:Kolm}, Matthaeus and Zhou
\cite{MattZhou}, and Zhou and Matthaeus \cite{ZhouMatt} proposed an
alternate phenomenology in which the energy spectra are proportional
to $k^{-5/3}$, similar to Kolmogorov's spectrum for fluid
turbulence. Current numerical \cite{MKV:mhdsim,Bisk1:Kolm,Bisk2:Kolm}
and theoretical \cite{MKV:B0RG,Srid1,Srid2} work support
Kolmogorov-like phenomenology for MHD turbulence. In the present paper
we show that Kolmogorov's spectrum ($\propto k^{-5/3}$) is a
consistent solution of renormalization group (RG) equation of MHD
turbulence.

Forster et al., DeDominicis and Martin, Fournier and Frisch, Yakhot
and Orszag \cite{fluidRG} applied RG technique to fluid turbulence.
They considered external forcing and calculated renormalized
parameters: viscosity, noise coefficient, and vertex.  McComb and his
coworkers \cite{McCoRG} instead applied a self-consistent RG
procedure; here the energy spectrum was assumed to be Kolmogorov's
powerlaw, and the renormalized viscosity was computed iteratively.
For MHD turbulence, Fournier et al., Camargo and Tasso, and Liang and
Diamond \cite{MHDRG} employed RG technique in the
similar lines as Forster et al.~\cite{fluidRG}.  In this letter we 
will apply McComb's self-consistent technique to MHD turbulence.
Earlier Verma \cite{MKV:B0RG} had done a self-consistent
calculation and showed that the mean magnetic field gets renormalized,
and the Kolmogorov's powerlaw is a consistent solution of MHD RG equation.
In this letter we will carry out the renormalization of viscosity
and resistivity.

For simplicity of the calculation we assume that the mean magnetic
field is absent.  This allows us to assume the turbulence to be
isotropic to a reasonable approximation.  In presence of mean magnetic
field, turbulence become anisotropic; this issue has been studied by
Sridhar and Goldreich \cite{Srid1} and Goldreich and Sridhar
\cite{Srid2}.  In addition to the above assumption, we also take cross
helicity $(2 {\bf u \cdot b})$, magnetic helicity $({\bf a \cdot b})$,
and kinetic helicity $({\bf u \cdot \omega})$ to be zero, where ${\bf
a}$ is magnetic vector potential, and ${\bf \omega}$ is the vorticity.

In our RG procedure the wavenumber range $(k_N,k_0)$ is divided
logarithmically into $N$ shells. We carry out the elimination of the
first shell $k^>=(k_1,k_0)$ and obtain the modified MHD equation for
$k^<=(k_{N},k_1)$.  This process is continued for all the shells. The
shell elimination is performed by ensemble averaging over $k^>$ modes
\cite{fluidRG,MHDRG}.  We assume that $u_i^>(\hat{k})$, and
$b_i^{>}(\hat{k})$ have
gaussian distributions with zero mean, while $u_i^{<}(\hat{k})$ and
$b_i^{<}(\hat{k})$ are unaffected by the averaging process.  In
addition we take
\begin{eqnarray}
\left\langle u_i^> (\hat{p}) u_j^> (\hat{q})\right\rangle  & = &
P_{ij}({\bf p)} C^{uu} (\hat{p}) \delta(\hat{p}+\hat{q}) 
\label{eqn:avgbegin}\\
\left\langle b_i^> (\hat{p}) b_j^> (\hat{q})\right\rangle  &= &
P_{ij}({\bf p)} C^{bb} (\hat{p}) \delta(\hat{p}+\hat{q}) \label{eqn:avgend}
\end{eqnarray}

Let us denote $\nu_{(n)}$ and $\lambda_{(n)}$ as the viscosity and
resistivity after the elimination of the $n$ shell.  To first order
of perturbation, we obtain
\bea 
\left( -i\omega + \nu_{(n)} k^2 + 
	\delta \nu_{(n)} k^2 \right) u_i^<(\hat{k}) & = & 
	-\frac{i}{2} P^+_{ijm}({\bf k}) \int d\hat{p}
	[u_j^< (\hat{p}) u_m^< (\hat{k}-\hat{p}) \nonumber \\ 
& & \hspace{1in} - b_j^<(\hat{p}) b_m^< (\hat{k}-\hat{p}) ] \\
\left( -i\omega + \lambda_{(n)} k^2 
	+ \delta \lambda_{(n)} k^2 \right) b_i^<(\hat{k}) & = &  
	-i P^-_{ijm}({\bf k}) \int d\hat{p} 
	[u_j^< (\hat{p}) b_m^< (\hat{k}-\hat{p}) ] 
\eea 
where
\bea 
\delta \nu_{(n)}(k) & = & \frac{1}{(d-1)k^2} 
			  \int^{\Delta}_{\hat{p}+\hat{q}=\hat{k}}
				 \frac{d {\bf p}}{(2 \pi)^d}
 			  [ S(k,p,q) \frac{C^{uu}(q)}
			  {\nu_{(n)}(p) p^2+\nu_{(n)}(q) q^2} 
			       \nonumber \\   
 		    & & \hspace{1.5in} - S_6(k,p,q) \frac{C^{bb}(q)}
			{\lambda_{(n)}(p) p^2+\lambda_{(n)}(q) q^2} ]
                         \label{eqn:nu} \\
\delta \lambda_{(n)}(k) & = & \frac{1}{(d-1)k^2}
			      \int^{\Delta}_{\hat{p}+\hat{q}=\hat{k}}
				 \frac{d {\bf p}}{(2 \pi)^d}
			      [ - S_8(k,p,q) \frac{C^{bb}(q)}
			      {\nu_{(n)}(p) p^2+\lambda_{(n)}(q) q^2}
				\nonumber \\   
                        &  & \hspace{1.5in} +S_9(k,p,q) \frac{C^{uu}(q)}
			     {\lambda_{(n)}(p) p^2+\nu_{(n)}(q) q^2} ]
                             \label{eqn:lambda}
\eea
with $S_i(k,p,q)$s as functions of $k,p$, and $q$.  Hence, after the
elimination of the $(n+1)$th shell, the effective viscosity and
resistivity will be $(\nu,\lambda)_{(n+1)} (k)=(\nu,\lambda)_{(n)} (k)
+\delta (\nu,\lambda)_{(n)} (k)$.

\begin{figure}
\centerline{
        \psfig{figure=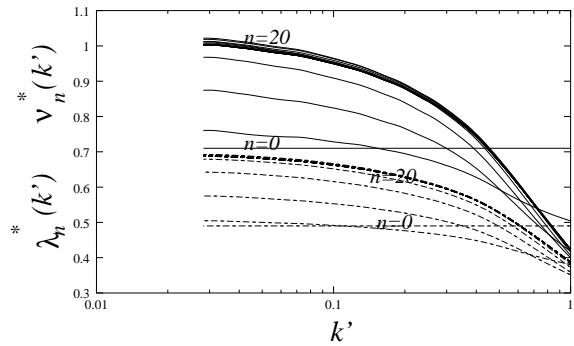,width=12cm,angle=0}}
        \vspace*{0.5cm}
\caption{Plot of $\nu^*(k')$ (solid lines) and $\lambda^*(k')$ (dashed
lines) vs. $k'$ for
$d=3$ and $\sigma_c=0, r_A=1$. Values at various iterations are
shown by different curves.}
\label{fig:d3ra1}
\end{figure}

We solve the above equations iteratively.  To simplify,
we substitute $C(k)$ in the Eqs.~(\ref{eqn:nu}, \ref{eqn:lambda}) 
by one dimensional energy spectrum $E(k)$
\begin{eqnarray}
C^{uu,bb}(k) = \frac{2 (2 \pi)^d}{S_d (d-1)} k^{-(d-1)} E^{u,b}(k) 
\end{eqnarray}
where $S_d$ is the surface area of $d$ dimensional spheres.  We assume
that $E^u(k)$ and $E^b(k)$ follow 
\be
	E^u(k)  =  K^u \Pi^{2/3} k^{-5/3}; 	\hspace{1cm}
	E^b(k) =  E^u(k) / r_A \label{eqn:Ek} 
\ee
Regarding $\nu_{(n)}$ and $\lambda_{(n)}$, we attempt the following
form of solution
\bea
(\nu,\lambda)_{(n)} (k_n k') & = & (K^u)^{1/2} \Pi^{1/3} k_n^{-4/3} 
(\nu,\lambda)_{(n)}^* (k') 
\eea

with $k=k_{n+1}k' (k' < 1)$ with the expectation that
$\nu_{(n)}^*(k')$ and $\lambda_{(n)}^*(k')$ are universal functions
for large $n$.  We numerically solve for $\nu_{(n)}^*(k')$ and
$\lambda_{(n)}^*(k')$.  Our calculations reveal that the solutions of
$\nu_{(n)}^*(k')$ and $\lambda_{(n)}^*(k')$ converge for all $d > d_c
\approx 2.2$.  From this observation we can conclude that Kolmogorov's
energy spectrum ($E(k) \propto k^{-5/3}$) is a consistent solution of
the RG equations.  Meanwhile, Kraichnan's $3/2$ energy spectrum and
$\nu k^2 = \lambda k^2 \propto k B_0$, where $B_0$ (a constant) is the
magnetic field of the large eddies, do not satisfy the renormalization
group equations [Eqs.~(\ref{eqn:nu}, \ref{eqn:lambda})].  Hence $E(k)
\propto k^{-3/2}$ is not a consistent solution of RG equations.  Our
result regarding the nonexistence of stable RG fixed point for $d=2$
is consistent with the RG calculation of Liang and Diamond
\cite{MHDRG}.  Refer to Fig.~(\ref{fig:d3ra1}) for illustration of
$\nu_{(n)}^*(k')$ and $\lambda_{(n)}^*(k')$ for $d=3$ and $r_A=1$.
 
The values of renormalized parameters for $d=3$ and various $r_A$ are
shown in Table \ref{tab:d3}. For large $r_A$, the asymptotic $\nu^*$
is close to the corresponding value for fluid turbulence, but the
asymptotic $\lambda^*$ is also comparable to $\nu*$.  
This implies that in fluid
dominated regime, there is a significant magnetic energy flux in addition
to the usual Kolmogorov's flux in fluid modes.  As $r_A$ is decreased,
$\nu^*$ increases but $\lambda^*$ decreases.  This trend is seen till
$r_A \approx 0.25$, where the RG fixed point with nonzero $\nu^*$ and
$\lambda^*$ becomes unstable, and the trivial RG fixed point with
$\nu^*=\lambda^*=0$ becomes stable.  This result suggests an absence
of turbulence for $r_A$ below 0.25. This is consistent with the fact
that MHD equations become linear in the $r_A \rightarrow 0$
(fully magnetic) limit.

We can proceed further and compute various cascade rates and 
Kolmogorov's constant for MHD using the renormalized parameters
computed above.  To compute these quantities we resort to the
energy equations, which are \cite{Stan,Dar:flux}
\bea
\left(\frac{\partial}{\partial t}  + 2 \nu k^2 \right) C^{uu}({\bf k},t) 
& = &	\frac{1}{(d-1)(2 \pi)^d \delta({\bf k+k'})} 
	\int_{\bf k'+p+q=0} \frac{d {\bf p}}{(2 \pi)^d} 
	       [S^{uu}({\bf k'|p|q})+S^{uu}({\bf k'|q|p}) \nonumber \\
&   & 	\hspace{2.5in}+S^{ub}({\bf k'|p|q})+S^{ub}({\bf k'|q|p})] \\
\left(\frac{\partial}{\partial t}  + 2 \lambda k^2 \right) C^{bb}({\bf k},t) 
& = &	\frac{1}{(d-1)(2 \pi)^d \delta({\bf k+k'})} 
	\int_{\bf k'+p+q=0} \frac{d {\bf p}}{(2 \pi)^d} 
	       [S^{bu}({\bf k'|p|q})+S^{bu}({\bf k'|q|p}) \nonumber \\
&   & 	\hspace{2.5in}+S^{bb}({\bf k'|p|q})+S^{bb}({\bf k'|q|p})]
\eea
where
\begin{equation}
S^{uu}({\bf k'|p|q}) = -\Im \left({\bf \left[k'.u(q)\right]} %
{\bf \left[u(k').u(p)\right]} \right) ,
\label{eq:Sukup_def}
\end{equation}    
\begin{equation}
S^{bb}({\bf k'|p|q}) = -\Im \left({\bf \left[k'.u(q)\right]} %
{\bf \left[b(k').b(p)\right]} \right) ,
\label{eq:Sbkbp_def}
\end{equation}
\begin{equation}
S^{ub}({\bf k'|p|q}) =  \Im \left({\bf \left[k'.b(q)\right]} %
{\bf \left[u(k').b(p)\right]} \right) ,
\label{eq:Sukbp_def}
\end{equation}
\begin{equation}
S^{bu}({\bf k'|p|q}) = - S^{ub}({\bf p|k'|q})  
\label{eq:Sbkup_def}
\end{equation}
Here $\Im$ stands for the imaginary part of the argument, and 
the above integrals have constraints that ${\bf k'+p+q = 0}$ (${\bf
k=-k'}$).  The energy equations in the above form have been
written by Dar et al. \cite{Dar:flux}, who interpret
the terms $S({\bf k|p|q})$
as energy transfer rate from mode ${\bf p}$ (second argument of
$S$) to ${\bf k}$ (first argument of $S$) with mode ${\bf q}$ (third
argument of $S$) acting as a mediator. This interpretation
of energy transfer due to Dar et al.~\cite{Dar:flux} is consistent
with the earlier formalism.  

We can derive an expression for the energy transfer rate or energy flux
from a wavenumber sphere using $S({\bf k'|p|q})$.  The formula for
the energy flux from inside of the $X$-sphere ($X<$) 
to outside of the $Y$-sphere ($Y>$) is
\bea 
\Pi^{X<}_{Y>}(k_0) & = & \int_{k>k_0} \frac{d {\bf k}}{(2 \pi)^d} 
		       \int_{p<k_0} \frac{d {\bf p}}{(2 \pi)^d}  
			\la S^{YX}({\bf k'|p|q}) \ra
\label{eqn:flux}		
\eea 
where $X$ and $Y$ stand for $u$ or $b$. In our study we assume
that the kinetic energy is forced at small wavenumbers, and the
turbulence is steady.  We calculate the above fluxes analytically
to the leading order in perturbation series using  the same
procedure as that of Leslie \cite{Lesl}.  The flux is calculated using
Eq.~(\ref{eqn:flux}) by taking ensemble average of $S^{YX}$.  The 
expression for $\la S^{bb}(k|p|q) \ra$ is
\bea
\la S^{bb}(k|p|q) \ra & = & \int_{-\infty}^t  dt'  [
                T_{4}(k,p,q) G^{bb}(k,t-t') C^{bb}(p,t-t') C^{uu}(q,t-t')
			\nonumber \\
& &\hspace{1cm} + T_{8}(k,p,q) G^{bb}(p,t-t') C^{bb}(k,t-t') C^{uu}(q,t-t')
			\nonumber \\
& &\hspace{1cm} + T_{10}(k,p,q) G^{uu}(q,t-t') C^{bb}(k,t-t') C^{bb}(p,t-t') ]
\eea
where $T_i (k,p,q)$ are functions of wavevectors $k,p$, and $q$.  The
expressions for other transfer rates $\la S^{uu}(k|p|q) \ra$, $\la
S^{ub}(k|p|q) \ra$, and $\la S^{bu}(k|p|q) \ra$ look similar.  In the
above formulas we substitute Kolmogorov's spectrum
[Eqs.~(\ref{eqn:Ek})] for the energy spectrum, and the
following expression for the effective viscosity and resistivity
\bea
(\nu,\lambda)(k) & = & (K^u)^{1/2} \Pi^{1/3} k^{-4/3} (\nu^*,\lambda^*)    
	\mbox{\hspace{1cm} for $k \ge k_n$} \label{eqn:nuk} 
\eea

Following the same procedure as Leslie \cite{Lesl}, we obtain the
following nondimensional form of the equations
\be
\frac{\Pi^{X<}_{Y>}}{\Pi} =  \frac{4 S_{d-1}}{(d-1)^2 S_d}
              (K^u)^{3/2} \int_0^1 dv \ln{(1/v)} \int_{1-v}^{1+v} dw 
		(vw)^{d-2} (\sin \alpha)^{d-3}
		F^{X<}_{Y>} 
\label{eqn:piub}
\ee
where the integrals $F^{X<}_{Y>}$ are function of $v$, $w$, $\nu^*$,
and $\lambda^*$.  After a bit of manipulation we can obtain
$\Pi^{X<}_{Y>}/\Pi$ and the constant $K^u$.  In addition we
can also obtain the Kolmogrov's constant $K$ for total energy
\be
E(k) = K \Pi^{2/3} k^{-5/3}
\ee
using $K = K^u (1+r_A^{-1}) $.  The values of $\Pi^{X<}_{Y>}/\Pi$
and $K$ for $d=3$ and various $r_A$ are listed in Table~\ref{tab:d3}.

The entries in the Table~\ref{tab:d3}
 show that the cascade rates \piulbg, \piblbg,
\piulbl, \piblug are approximately of the same order for $r_A$
between 0.5 and 1, but the flux \piulug is rather small.  The sign of
\piblbg is positive, indicating that the ME cascades forward, that is
from large length-scales to small length-scales.  The magnetic energy
thus appearing at small length-scales will be lost due to resistive
dissipation, and the large-scale magnetic field is maintained by the
\piulbl flux.  The Kolmogorov's constant $K$ is approximately constant
and is close to 1.6, same as that for fluid turbulence ($r_A=\infty$),
all $r_A$ greater than 0.5.

To summarize, we employed a self-consistent RG scheme to MHD 
turbulence and found that Kolmogorov's 5/3 powerlaw is a consistent
solution of RG equations for $d \ge d_c \approx 2.2$.  For Kolmogorov's
solution, the renormalized viscosity and resistivity have been
calculated, and they are found to be positive.  For $d=3$,
variation of $\nu^*$ and $\lambda^*$ with $r_A$ shows some
interesting features.  For large $r_A$, $\nu^*$ is same as that
for fluid turbulence, but $\lambda^*$ is also nonzero, in fact larger
than $\nu^*$.  As $r_A$ is decreased, $\nu^*$ increases but 
$\lambda^*$ decreases until $r_A \approx 0.25$ at which value
turbulence disappears.

Using the flux equations we have obtained various fluxes and
Kolmogorov's constant $K$.  For $d=3$, $K$ does not vary significantly
with the variation of $r_A$, and it is close to $K$ for fluid
turbulence.  We find that the cascade rate from magnetic-sphere to
outside magnetic-sphere ($\Pi^{b<}_{b>}$) is positive, a result consistent
with the numerical results of Dar et al. \cite{Dar:flux}.

In this paper we have restricted ourselves to nonhelical turbulence.
Helical MHD turbulence is very important specially in the light
of enhancement of magnetic energy (dynamo).  However, 
the physics of helical turbulence is more complex with the
appearance of inverse cascade of magnetic helicity etc.
The field-theoretic analysis for this case will be taken up
later.  Recent studies show that the  mean magnetic field  has
a strong effect on energy spectrum, and it induces anisotropy.
 A full-fledge field theory calculation
in the presence of mean magnetic field is also necessary for 
a clearer picture of MD turbulence

The author thanks J. K. Bhattacharjee for very valuable discussions
and ideas.  He also thank G. Dar and V. Eswaran for many insights
from numerical results. 


\begin{thebibliography}{10}

\bibitem{Krai65}
R.~H. Kraichnan, Phys. Fluids {\bf 8},  1385  (1965).

\bibitem{Iros}
P.~S. Iroshnikov, Sov. Astron. I. {\bf 7},  566  (1964).

\bibitem{Mars:Kolm}
E. Marsch,  in {\em Reviews in Modern Astronomy}, edited by G. Klare
  (Springer-Verlog, Berlin, 1990), p.\ 43.

\bibitem{MattZhou}
W.~H. Matthaeus and Y. Zhou, Phys. Fluids B {\bf 1},  1929  (1989).

\bibitem{ZhouMatt}
Y. Zhou and W.~H. Matthaeus, J. Geophys. Res. {\bf 95},  10291  (1990).

\bibitem{MKV:mhdsim}
M.~K. Verma {\it et~al.}, J. Geophys. Res. {\bf 101},  21619  (1996).

\bibitem{Bisk1:Kolm}
W.~C. M{\"{u}}ller and D. Biskamp, Phys. Rev. Lett. {\bf 84},  475  (2000).

\bibitem{Bisk2:Kolm}
D. Biskamp and W.~C. M{\"{u}}ller, Phys. Plasma {\bf 7},  4889  (2000).

\bibitem{MKV:B0RG}
M.~K. Verma, Phys. Plasma {\bf 6},  1455  (1999).

\bibitem{Srid1}
S. Sridhar and P. Goldreich, Astrophys. J. {\bf 432},  612  (1994).

\bibitem{Srid2}
P. Goldreich and S. Sridhar, Astrophys. J. {\bf 438},  763  (1995).

\bibitem{fluidRG}
D. Forster, D.~R. Nelson, and M.~J. stephen, Phys. Rev. A {\bf 16},  732
  (1977);
C. DeDominicis and P.~C. Martin, Phys. Rev. A {\bf 19},  419  (1979);
J.~D. Fournier and U. Frisch, Phys. Rev. A {\bf 17},  747  (1979);
V. Yakhot and S.~A. Orszag, J. Sci. Comput. {\bf 1},  3  (1986).

\bibitem{McCoRG}
W.~D. McComb, {\em The Physics of Fluid Turbulence} (Claredon, Oxford, 1990);
W.~D. McComb, Rep. Prog. Phys. {\bf 58},  1117  (1995).

\bibitem{MHDRG}
J.~D. Fournier, P.-L. Sulem, and A. Pouquet, J. Phys. A {\bf 15},  1393
  (1982);
S.~J. Camargo and H. Tasso, Phys. Fluids B {\bf 4},  1199  (1992);
W.~Z. Liang and P.~H. Diamond, Phys. Fluids B {\bf 5},  63  (1993).

\bibitem{Stan}
M.~M. Stani{\u{s}}i{\'{c}}, {\em Mathematic Theory of Turbulence}
  (Springer-Verlog, New York, 1988).

\bibitem{Dar:flux}
G. Dar, M.~K. Verma, and V. Eswaran, Physica D, {\bf 157}, 207, 2001.

\bibitem{Lesl}
D.~C. Leslie, {\em Development in the Theory of Turbulence} (Claredon, Oxford,
  1973).

\end{thebibliography}

\begin{table}
\caption{The values of $\nu^*,\lambda^*,\nu^{uu*},\nu^{ub*},
\lambda^{bu*},\lambda^{bb*}$ for for various $r_A$ when ${\bf d=3}$
and $\sigma_c=0$}
\label{tab:d3}
\begin{center} 
\begin{tabular}{lcccccrcc} 
$r_A$  & $\nu^*$ & $\lambda^*$ & \piulug/$\Pi$ & \piulbg/$\Pi$ &
\piblug/$\Pi$ & \piblbg/$\Pi$  & \piulbl/$\Pi$ & K \\ \hline

$\infty$ & 0.38 & $--$ & 1 & $--$ & $--$ & $--$ & $--$ & 1.53 \\
5000 & 0.36 & 0.85 & 1    & 3.5E-4 &-1.05E-4 & 2.4E-4 & 1.3E-4 & 1.51 \\
5    & 0.47 & 0.82 & 0.61 & 0.26 & -0.050 & 0.19 & 0.13 & 1.51 \\
1    & 1.00 & 0.69 & 0.12 & 0.39 &  0.12  & 0.37 & 0.49 & 1.50  \\

0.5  & 2.11 & 0.50 & 0.037 & 0.33 & 0.33  & 0.30 & 0.63 & 1.65 \\
0.3  & 11.0 & 0.14 & 0.011 & 0.36 & 0.42  & 0.21 & 0.63 & 3.26 \\
0.2  & $--$ & $--$ & $--$  & $--$ & $--$  & $--$ & $--$ & $--$ \\
\end{tabular}
\end{center}
\end{table}

\end{document}